\documentclass[a4paper,11pt]{article}
\usepackage{jinstpub} 
\usepackage{lineno}

\title{\boldmath Production of GEM-like structures for cryogenic applications, using laser-cutting techniques}

\author[a,b,1]{D. Rodas-Rodríguez,\note{Corresponding author.}}
\author[a]{A.\,F.\,V.~Cortez,}
\author[a]{M. Ku\'zniak,}
\author[b]{D. González-Díaz,}
\author[a]{P.\,A.\,O.\,C.~Silva,}
\author[a]{A.~Gnat,}
\author[a]{G. Nieradka,}
\author[a]{T. Sworobowicz,}
\author[b]{E. Alario,}
\author[c]{C.\,D.\,R.~Azevedo,}
\author[d]{K.\,T.~Floethner,}
\author[e,f]{P. Gasik,}
\author[b]{J. Llerena,}
\author[g]{C.\,M.\,B.~Monteiro,}
\author[d]{R.~Oliveira,}
\author[b]{A.~Pallas,}
\author[b]{D.~Tenreiro,}
\author[d]{V. Peskov,}
\author[g]{J.\,M.\,F. dos Santos}

\affiliation[a]{Astrocent, Nicolaus Copernicus Astronomical Center of the Polish Academy of Sciences, \\Rektorska 4, 00-614 Warsaw, Poland}
\affiliation[b]{Universidade de Santiago de Compostela,\\Campus Universitario Sur, ES-15782, Spain}
\affiliation[c]{Institute of Nanostructures, Nanomodelling and Nanofabrication (i3N), Universidade de Aveiro,\\Campus Universitario de Santiago, 3810-193 Aveiro, Portugal}
\affiliation[d]{CERN,\\Esplanade des Particules 1, Meyrin, Switzerland}
\affiliation[e]{GSI Helmholtzzentrum für Schwerionenforschung GmbH (GSI),\\Darmstadt, Germany}
\affiliation[f]{Facility for Antiproton and Ion Research in Europe GmbH (FAIR),\\Darmstadt, Germany}
\affiliation[g]{LIBPhys-UC, Department of Physics, University of Coimbra,\\Rua Larga, 3000 Coimbra, Portugal}
\emailAdd{drodas@camk.edu.pl}

\abstract{A novel concept for electroluminescence (EL) structures was recently proposed. In it, a wavelength-shifting material is deposited inside the holes of GEM-like structures which, after suitable optical treatment of its electrodes, improves the light collection and detection efficiency in noble gas TPCs. This new development directly addresses problems related with the scalability of future dual-phase TPCs for rare-event searches, matching (and potentially exceeding) the performance of conventional EL techniques.

We report the newest developments on the production of such structures using laser-based techniques, namely the manufacture of a first batch of the so-called FAT-GEMs. This process allows low-cost and reproducible manufacturing of a high volume of such structures.

In addition to the detailed description of the production, we present a performance assessment in pure argon, at a gas density close to the one expected in LAr conditions. An energy resolution of 23.5$\pm$1~\% (FWHM) at 5.9~keV was obtained, indicating a consistent improvement over previous batch. The optical treatment of the electrode surfaces has been greatly simplified and modestly improved, while charging-up effects arising from the use of laminates eliminated.}


\begin{document}
\maketitle
\flushbottom

\section{Introduction}
\label{sec:intro}
Dual-phase Time Projection Chambers (TPCs) are widely used in rare-event searches for direct dark matter (DM) detection and neutrino physics, thanks to their low energy thresholds and excellent background discrimination capabilities~\cite{LZ, XENONnT, DarkSide50}. Relevant particle interactions are expected to release an energy in the liquid target of O(10)~keV, resulting in a prompt scintillation signal (S1) accompanied by primary ionization. The ionization electrons are drifted to the liquid–gas interface and extracted into the gas phase, where electroluminescence (EL) generates a secondary scintillation signal (S2) recorded by photosensors~\cite{DarkSide20k_2, aprilenoblegas}. Combining the information from these signals, experiments such as LZ~\cite{LZ}, XENONnT~\cite{XENONnT}, and DarkSide-50~\cite{DarkSide50} have set the most stringent sensitivity limits on Weakly Interacting Massive Particles (WIMPs), to date. These results motivate the development of next-generation detectors with target masses of several tens of tonnes~\cite{DarkSide20k}. 

However, scaling EL-based detectors to large areas introduces significant challenges. Noble elements scintillate predominantly in the vacuum ultraviolet (VUV), where the photon detection efficiency of most photosensors is limited, particularly for argon at 128~nm~\cite{BoniventoArTech}. In addition, defining a stable and uniform EL region over square-meter scales is technically demanding. The current use of parallel meshes or wire planes~\cite{MonrabalNextWhite}, while offering good optical transparency and energy resolution, becomes increasingly problematic due to mechanical deformation, electrostatic sagging, leading to electric field non-uniformities, that affect the detector performance.

To address these limitations, novel optical amplification structures have been proposed~\cite{CortesiTHGEM, Ban2017, MavrokoridisGlassGEM}. Among them, Field Assisted Transparent Gaseous Electroluminescence Multipliers (FAT-GEMs)~\cite{gonzalez2020new} offer a robust, low-radioactivity, and scalable alternative to traditional meshes, achieving comparable energy resolution~\cite{leardini2024fat}. These thick, transparent PMMA-based structures are intrinsically radiopure, compatible with cryogenic operation, and have been successfully coupled to solid wavelength shifters, making them particularly attractive for argon-based detectors~\cite{kuzniak2021development}.

In this work, we present the manufacturing process of an enhanced FAT-GEM design developed at the Astrocent facilities using laser-based clean cutting techniques. The new design is optimized to enhance light collection efficiency in argon detectors while preserving the key requirements for rare-event searches.

\section{Field Assisted Transparent Gaseous Electroluminescence Multiplier (FAT-GEM)}
\label{sec:FAT-GEM}

\begin{figure}[h!]
    \centering
    \includegraphics[width=0.5\linewidth]{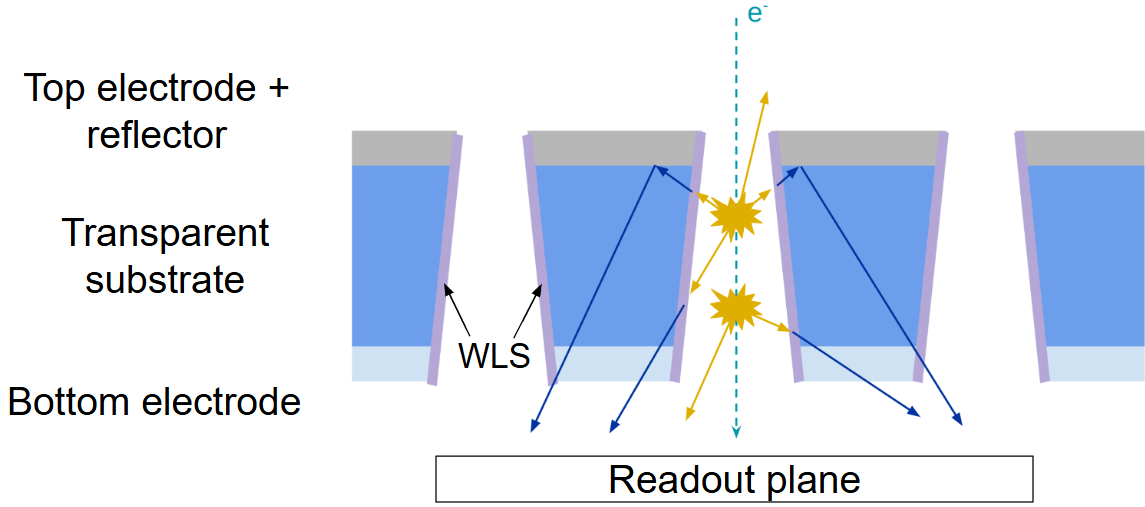}
    \caption{Schematic representation of a typical FAT-GEM configuration with a combination of reflective and transparent electrodes, with a transparent substrate and WLS material deposited in the conical-shaped holes achieved through laser drilling. The scheme also represents how the light wavelength is shifted from the VUV emmission (yellow) to the visible range (blue).}
    \label{fig:fatgem-scheme}
\end{figure}

FAT-GEMs are structures based on Micro Pattern Gaseous Detector (MPGD) technology and optimized for optical readout, specifically developed for EL-based operation in noble-element TPCs \cite{leardini2024fat}. These thick structures, in the range of several millimetres, are significantly thicker than conventional GEM-like devices (Gas Electron Multiplier structures), which typically range from tens of microns to sub-millimetre scale (th-GEMs). This thickness enables a larger EL-gap to be defined thus enhancing the EL yield without reaching the charge multiplication regime, maintaining mechanical stability free from sagging and other potential deformation mechanisms.
Figure~\ref{fig:fatgem-scheme} shows a typical configuration of these structures. 

At its current stage of development, the manufacturing process of FAT-GEMs is highly customizable, allowing the structures to be tailored to the requirements of different applications. For instance, electrodes can be made optically transparent using materials such as indium tin oxide (ITO) or conductive polymers (e.g. PEDOT:PSS), gaining sensitivity to primary scintillation light (S1-enhanced configurations). Alternatively, combinations of transparent and reflective electrodes can be employed to increase light collection from secondary signals (S2-enhanced configurations). Similarly, the structure thickness can be varied to control the EL-gap, and different hole patterns (e.g. square or hexagonal) or hole shapes (cylindrical or conical) can be selected. More generally, the choice of bulk or electrode materials, surface treatments, coatings, and other aspects can be adapted to optimize the detector response (i.e. enhancing S1 or S2 detection, correction of optical effects, minimization of optical positive feedback, etc.) at different operational conditions and for distinct physics goals. In this proceeding we discuss those adaptations of direct relevance to dual-phase use in argon-based chambers.

\section{Fabrication process}

Previous productions of FAT-GEM structures have demonstrated performance comparable to that of parallel meshes in terms of energy resolution in gaseous Xenon \cite{leardini2024fat}, while also highlighting limitations related to the fabrication process. The current upgrade focuses on developing clean and scalable manufacturing techniques, specifically targetting at operation in argon, in particular at cryogenic conditions.

\subsection*{Material selection}
Poly(methyl methacrylate) (PMMA) is used as the bulk material due to its radiopurity, high optical transparency in the visible range, and well-established mechanical properties. To make PMMA VUV transparent, wavelength-shifting (WLS) coatings are used. It has been previously demonstrated that coating the hole walls with TPB converts VUV photons into visible light, with an emission peak around 420~nm \cite{DarkSide50TPB}, enabling efficient light collection with standard photosensors.

Previous batches of FAT-GEMs employed ITO adhesive films as transparent electrodes on both faces of the substrate. In addition, an Enhanced Specular Reflector (ESR) film was fixed using a transparent adhesive to one face in order to improve light collection. Despite the excellent detector performance achieved with this configuration, several drawbacks were identified, including long-term charging-up effects (at the scale of minutes) at the ITO-ESR-substrate interfaces due to the presence of multiple dielectric layers, affecting light transport and collection uniformity. 

The current production adopts a revised electrode configuration, using a magnetron-sputtered ITO layer to maintain optical transparency on one face, while in the opposite face Aluminum is deposited, which serves simultaneously as an electrode and a reflector. Compared to previous productions, the optical transparency of the ITO layer has been enhanced by approximately 10\%, further improving light transmission. This approach also reduces the amount of dielectric material in the structure, mitigating charging-up effects while preserving high reflectivity in the visible range and improving the overall optical performance.

\subsection*{FAT-GEM production}
\begin{figure}[h]
\centering
\begin{minipage}{0.4\textwidth}
  \centering
  \includegraphics[width=\linewidth]{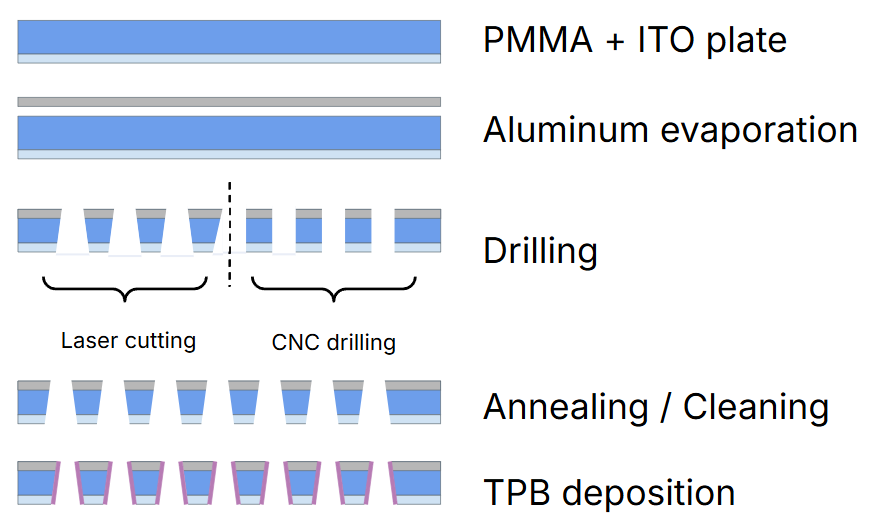}
\end{minipage}\hfill
\begin{minipage}{0.24\textwidth}
  \centering
  \includegraphics[width=0.97\linewidth]{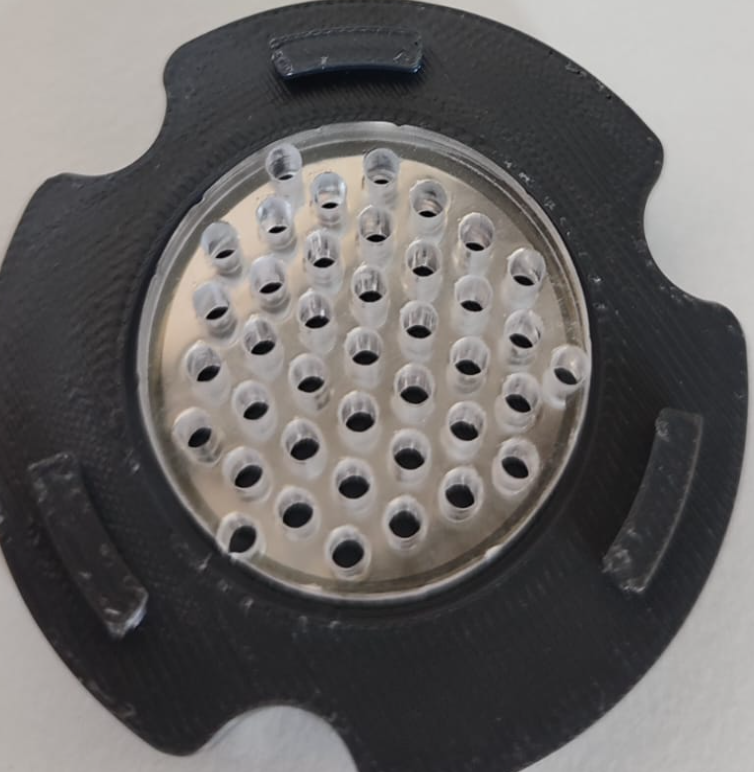}
\end{minipage}\hfill
\begin{minipage}{0.24\textwidth}
  \centering
  \includegraphics[width=\linewidth]{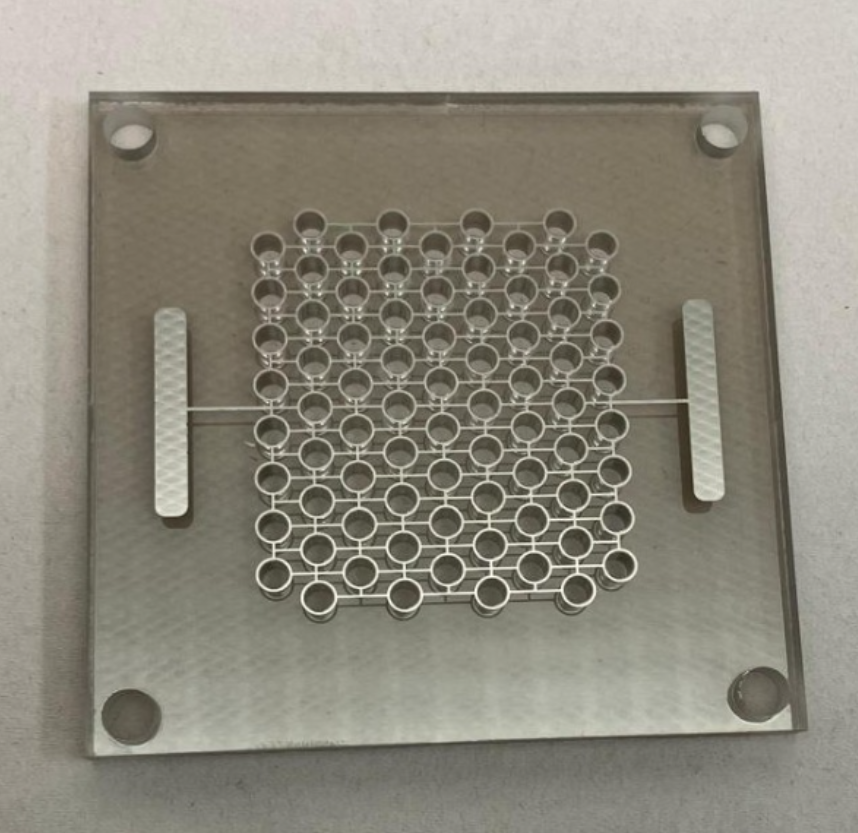}
\end{minipage}\hfill
\caption{Left: Scheme of the production steps for the upgraded and scalable FAT-GEM manufacturing.. Right: Finalized samples showing different configurations: S2-enhanced (left) and S1-enhanced (right), in this case with a VUV-transparent mesh-like cathode.}
\label{fig:production-scheme}
\end{figure}

The production of this initial batch was performed at Astrocent and CEZAMAT, Warsaw University of Technology (Poland)~\cite{kuzniak2021development, leardini2024fat}, with a second ongoing at CERN. Figure~\ref{fig:production-scheme} shows a schematic with all production steps along with typical FAT-GEM structures (produced at Astrocent and at CERN).

For the production of the FAT-GEM structures described in this work, we used 5~mm-thick hard-coated PMMA plates with a magnetron-sputtered ITO (indium tin oxide) coating of 400~$\Omega$/sq surface resistivity and 90\% transparency to 420~nm light according to the producer (Visiontek Systems Ltd.).
\\
\textbf{i) Sample preparation}. From the original plates (28$\times$28~cm$^2$) a total of six samples with 7$\times$7~cm$^2$ were cut using a CO$_2$ laser (xTool~P2), including the support holes in the corners. Before performing the aluminum deposition, samples were ultrasonically cleaned for 1~h in a solution of Alconox® in UPW (ultrapure water) and in UPW alone, and taken to the oven to dry (40 min at 78$^{{\circ}}$C). The edges of each tile were then covered using a stainless steel mask to ensure proper electrical insulation, eliminating the probability of surface discharges in that region.
\\
\textbf{ii) Aluminium deposition}. To perform the aluminum deposition, a vacuum evaporator (Korvus Hex Modular Deposition System) was used. The target thickness of the coatings, approximately 400~nm, was monitored using a thin-film thickness sensor. Before deposition, the samples were placed in a holder fixed to the top of the evaporator and aligned with the tungsten coil. Evaporation started only when the pressure inside the chamber was below 2$\times$10$^{-5}$~mbar. To prevent oxidation of the aluminum, the temperature inside the evaporator was controlled by adjusting the current injected into the coil, with the temperature being monitored by a temperature sensor placed inside the evaporator and in the vicinity of the crucible. 
After the aluminum deposition, the samples were visually inspected to assess the presence of irregularities, including Al-coating uniformity, possible transparency and reflectivity issues, as well as darkening effects originating from overheating during the process. Following that the surface conductivity was assessed using a multimeter (ANENG Q1).
\\
\textbf{iii) Hole pattern production (Laser cutting and Annealing)}. After these initial quality control checks are complete, the hole pattern was perforated onto the different samples using a CO$_2$ laser cutting station. During this process, the structures were placed on a custom 3D-printed holder that lifts them 6~mm above the metallic support frame, preventing back-reflections that could produce non-uniformities (resulting from melting). The laser parameters used were: a power range of 1–10\%, a cutting speed of 25~mm/s, and a single pass.
Once the hole-pattern is complete, the samples were annealed to minimize any residual stress introduced in the PMMA by the laser drilling, which could potentially jeopardize the mechanical integrity in cryogenic conditions. For the annealing stage, a programmable oven was used. The annealing was performed using a similar approach to the one used in \cite{acrylicstudies}. The heating rate used was 20$^{{\circ}}$C/h, annealing was performed during 5h at 80$^{{\circ}}$C, and a cooling rate of 15$^{{\circ}}$C/h was used to get the samples to room temperature.
\\
\textbf{iv) TPB deposition}.Two processes were developed: one optimized for coating the holes and another to the surface coating, which were performed separately. Before TPB deposition, the chamber was cleaned to ensure no contamination from residues originating from other materials, and the evaporation coil was replaced with an appropriate boat with TPB. The samples were placed in a custom-made support fixed to the top of the evaporator and aligned with the TPB evaporation boat. TPB was weighted and optimized to obtain a uniform deposition of about 1.5~micron-thick layer on the top surface, to optimize the WLS efficiency, similar to the one used for the hole coating. 
The evaporation was only initiated once the pressure inside the chamber was below 5$\times$10$^{-4}$~mbar. During the evaporation, the temperature of the boat was controlled by varying the current and ensuring that this would be stable around 200$^{{\circ}}$C. 
\\
\textbf{v) Cleaning and Packaging}. Following this last step, with the help of an air-brush any residues at the surface were removed, the electrodes' surface resistivity/conductivity was assessed, and the structures were visually inspected using a microscope. After these final tests, the samples were packed in custom-made transport opaque boxes, placed inside vacuum-sealed plastic bags with dessicant bags, and then transported to Santiago de Compostela (Spain), for the initial performance assessment.
Table~\ref{tab:fatgem-summary} summarizes the sample structures produced in the first batch and studied in this work.
\begin{table}[h]
\centering
\caption{Description of the produced samples. All samples are fabricated from 5~mm thick PMMA plates. The parameter $\Delta$ refers to the difference between the hole diameters on the two faces of the structure.\label{tab:fatgem-summary}. Characterization of structures C and E are presented in Section \ref{sec:characterization}.}
\resizebox{\textwidth}{!}{%
\begin{tabular}{|c|c|c|c|c|c|c|c|c|}
\hline
ID & Dimensions & Bulk & Electrodes & TPB & Size/Pitch (mm) & $\Delta$ (mm)  & Hole shape \\
\hline
A & 7$\times$7~cm$^2$ & PMMA & ITO + Al (mesh) & Holes + surface & 2 / 4 & - & Cylindrical\\
B & 7$\times$7~cm$^2$ & PMMA & ITO + Al & No & 2 / 4 & - & Cylindrical\\
C & 5$\times$5~cm$^2$ & PMMA & ITO + Al & Holes & 2 / 4 & 0.259~$\pm$~0.095 & Conical\\
D & 5$\times$5~cm$^2$ & PMMA & ITO + Al & No & 2 / 4 & 0.259~$\pm$~0.095 & Conical\\
E & 5$\times$5~cm$^2$ & PMMA & ITO + Al & Holes & 3 / 5 & - & Cylindrical\\
F & $\varnothing$ 3~cm & PMMA & ITO + Al & Holes & 2 / 4 & - & Cylindrical\\
G & $\varnothing$ 3~cm & PMMA & ITO + Al & Holes & 2 / 4 & 0.194~$\pm$~0.062 & Conical\\
H & $\varnothing$ 3~cm & PMMA & ITO + Al & No & 2 / 4 & - & Cylindrical\\
I & $\varnothing$ 3~cm & PMMA & ITO + Al & No & 2 / 4 & 0.194~$\pm$~0.062 & Conical\\
\hline
\end{tabular}
}
\end{table}

\section{Performance Assessment}
\label{sec:characterization}

\subsection*{Experimental Setup}
An existing setup, described in detail in~\cite{leardini2024fat}, was used. It is placed at the University of Santiago de Compostela, Spain, and was used to characterize the structures. It consists of a high-pressure gas vessel filled with Argon (purity $\times$6) equipped with a 5.9~keV X-ray source placed at the cathode, a Teflon adapter designed to hold the FAT-GEM under test, and a highly transparent mesh set to ground. This prevents fringe-fields leaking into the PMT vacuum and suitably stabilizes the field in the buffer region downstream of the FAT-GEM anode. The PMT was then connected to a pre-amplifier (ORTEC 142), an amplifier (ORTEC 572A), and finally to an MCA (Amptek 8000D) to collect the spectra. The vessel is connected to a pressure/vacuum system, and prior to filling, it is pumped to a reference pressure of approximately 10$^{-4}$~mbar. During operation, the gas is continuously circulated through a cold getter, removing residual impurities. 

\subsection*{Results and Discussion}

Before characterization in gaseous argon, the samples were tested in dry air to assess electrical stability. The 5~mm thick structures withstood voltage drops in the range of 10-12~kV before the onset of electrical discharges. It was observed that such discharges can damage the ITO coating, a 100~M$\Omega$ resistor was connected in series between the power supply and the electrodes to limit the discharge current, after which normal operation was recovered.

For each structure, the drift field was fixed at the optimal value previously reported for similar FAT-GEMs in~\cite{leardini2024fat}, while a scan in the EL field from 2 to 4.25~kV$\cdot$cm$^{-1}\cdot$bar$^{-1}$ was performed. The analysis follows the procedure described in~\cite{leardini2024fat}. Briefly, for each field setting, MCA spectra were recorded, rebinned, background-subtracted, and fitted to a Gaussian function.

\begin{figure}[h]
    \centering
    \includegraphics[width=0.85\linewidth]{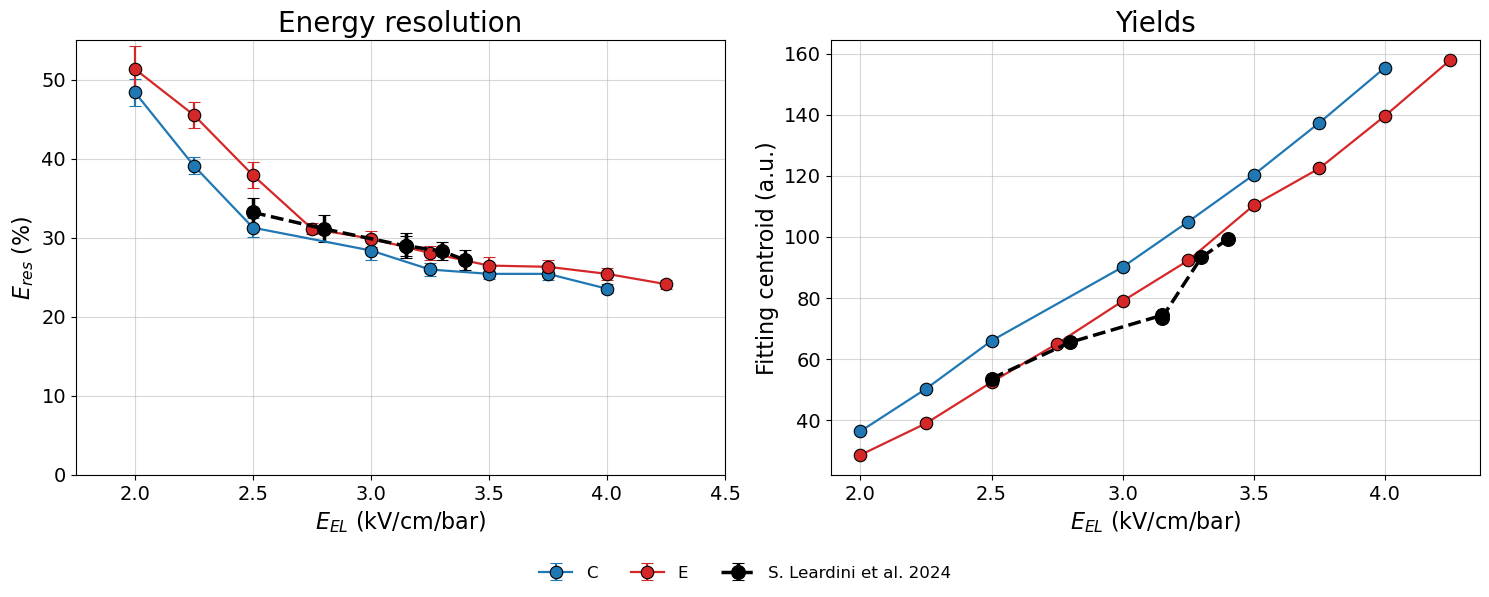}
    \caption{Left: Energy resolution (\%). Right: Scintillation yield. Results are shown for structures C and E in comparison with previous production \cite{leardini2024fat}.}
    \label{fig:results}
\end{figure}

Results from the quality assurance of the structures are summarized in Fig.~\ref{fig:results} for two configurations with different hole diameter-to-pitch configurations (C and E, see Table~\ref{tab:fatgem-summary}), measured at 4~bar (LAr-equivalent density) and compared with results from earlier productions~\cite{leardini2024fat}. As expected, the scintillation yield increases with the pressure-reduced EL field, accompanied by an improvement in energy resolution. Structures with 2~mm hole diameter show higher light yields than those with 3~mm holes. Overall, the upgraded structures show an average improvement in light yield of 21\% relative to previous productions~\cite{leardini2024fat}, loading to an energy resolution of 23.5$\pm$1.0\% (FWHM) at 5.9~keV for a pressure-reduced field of 4.0~kV$\cdot$cm$^{-1}\cdot$bar$^{-1}$ using structure~C (see Table~\ref{tab:fatgem-summary}). A more detailed study of the impact of hole size and geometry will be presented in a future publication.

In addition, stable operation was achieved without visible degradation in the performance, allowing operation in higher electroluminescence fields without discharges, up to 4.25~kV$\cdot$cm$^{-1}\cdot$bar$^{-1}$ at 4~bar, limited by the electrical insulation of the cathode. No evidence of charging-up effects was observed. Furthermore, one structure was intentionally subjected to electrical discharges to assess the robustness of the TPB coating, which remained unaffected after these tests.

\section{Conclusions}
\label{sec:conclusion}

A thorough review of the fabrication process of FAT-GEMs was undertaken at Astrocent facilities. The main modifications introduced in the production process were related with the electrode production and hole-pattern machining. In this new approach, magnetron sputtering was used for the deposition of ITO directly onto the PMMA, enhancing the optical transparency from 80\% (structures with the adhesive film) to about 90\%. Another relevant change was introduced in the production, replacing the ESR foils by deposited aluminum, which eliminated the charging up effects observed previously that were originated by the use of the adhesive film, while retaining a similar reflectivity. 
Furthermore, the introduction of a novel laser cutting technique for the production of the hole-pattern made possible obtaining conical holes instead of the traditional cylindrical shape ones. The implications of the different hole geometries are currently under investigation. 

The new FAT-GEMs were characterized in argon, at a density close to that of LAr conditions,  in our quality-assurance station located at Santiago de Compostela. They show a consistent improvement over previous structures, translating into a 20\% relative improvement of the light yield.

We reported as well the manufacturing of a second FAT-GEM batch at CERN (S1-enhanced, see Fig.~\ref{fig:production-scheme}-right) using CNC techniques and photo-litography, to produce a semitransparent-cathode FAT-GEM. The performance of these new structures and detailed comparisons with simulations will be reported reported in a follow-up paper.

\acknowledgments
This work was supported by the National Science Centre (NCN) SONATA-19 Project No.
2023/51/\linebreak[0]/D/ST2/02297 and from the European Union’s research and innovation programme Horizon Europe under the Marie Skłodowska-Curie grant agreement No 101154972. 

This research was also funded by the Spanish Ministry of Science and Innovation through the “Proyectos de Generación de Conocimiento” programme (PID2021-125028OB-C21 and PID2021-125028OB-C22). Additional financial support was provided by the Xunta de Galicia through the Centro Singular de Investigación de Galicia accreditation (2019–2022), and by the Spanish national “María de Maeztu” Units of Excellence programme (MDM-2016-0692).


\bibliographystyle{JHEP}
\bibliography{biblio}

@article{leardini2024fat,
  title={FAT-GEMs:(field assisted) transparent gaseous-electroluminescence multipliers},
  author={Leardini, S and Sa{\'a}-Hern{\'a}ndez, A and Ku{\'z}niak, M and Gonz{\'a}lez-D{\'\i}az, D and Azevedo, CDR and Lucas, F and Amedo, P and Cortez, AFV and Fern{\'a}ndez-Posada, D and Mehl, B and others},
  journal={Frontiers in Detector Science and Technology},
  volume={2},
  pages={1373235},
  year={2024},
  publisher={Frontiers Media SA}
}

@article{kuzniak2021development,
  title={Development of very-thick transparent GEMs with wavelength-shifting capability for noble element TPCs},
  author={Ku{\'z}niak, Marcin and Gonz{\'a}lez-D{\'\i}az, Diego and Amedo, Pablo and Azevedo, Carlos Davide Rocha and Fern{\'a}ndez-Posada, David Jos{\'e} and Ku{\'z}wa, Maciej and Leardini, Sara and Leonhardt, A and {\L}ȩcki, T and Manzanillas, Luis and others},
  journal={The European Physical Journal C},
  volume={81},
  number={7},
  pages={609},
  year={2021},
  publisher={Springer}
}

@inproceedings{gonzalez2020new,
  title={A new amplification structure for time projection chambers based on electroluminescence},
  author={Gonz{\'a}lez-D{\'\i}az, D and Fonta{\'\i}{\~n}a, M and Castro, D Garc{\'\i}a and Mehl, B and De Oliveira, R and Williams, S and Monrabal, F and Querol, M and {\'A}lvarez, V},
  booktitle={Journal of Physics: Conference Series},
  volume={1498},
  number={1},
  pages={012019},
  year={2020},
  organization={IOP Publishing}
}

@article{DarkSide20k,
  author = "{DarkSide Collaboration}",
  title = "{DarkSide-20k: A 20 tonne two-phase liquid argon time projection chamber for dark matter searches at LNGS}",
  journal = "Eur. Phys. J. Plus",
  volume = "133",
  pages = "131",
  year = "2018",
  eprint = "1707.08145",
  archivePrefix = "arXiv",
  primaryClass = "physics.ins-det"
}

@article{XENONnT,
  author = "{XENON Collaboration}",
  title = "{First Dark Matter Search Results from the XENONnT Experiment}",
  journal = "Phys. Rev. Lett.",
  volume = "131",
  pages = "041003",
  year = "2023",
  eprint = "2303.14729",
  archivePrefix = "arXiv",
  primaryClass = "hep-ex"
}

@article{LZ,
  author = "{LZ Collaboration}",
  title = "{First Dark Matter Search Results from the LUX-ZEPLIN (LZ) Experiment}",
  journal = "Phys. Rev. Lett.",
  volume = "131",
  pages = "041002",
  year = "2023",
  eprint = "2207.03764",
  archivePrefix = "arXiv",
  primaryClass = "hep-ex"
}

@article{DarkSide50,
  author       = {Agnes, P. and others},
  title        = {DarkSide-50 532-day Dark Matter Search with Low-Radioactivity Argon},
  journal      = {Phys. Rev. D},
  volume       = {98},
  pages        = {102006},
  year         = {2018},
  doi          = {10.1103/PhysRevD.98.102006},
}

@article{DarkSide50TPB,
  author       = {Agnes, P. and others},
  title        = {First results from the DarkSide-50 dark matter experiment at Laboratori Nazionali del Gran Sasso},
  journal      = {Physics Letters B},
  volume       = {743},
  pages        = {456–466},
  year         = {2015},
  doi          = {10.1016/j.physletb.2015.03.012},
}

@book{aprilenoblegas,
    author = {Aprile, E and Bolotnikov, A. and Bolozdynya, A. and Doke, T. },
    title = {Noble Gas Detectors},
    publisher = {Wiley YCH},
    year = {2007}
}

@article{DarkSide20k_2,
  author       = {Agnes, P. and others},
  title        = {DarkSide-20k sensitivity to light dark matter particles},
  journal      = {Comm. Phys.},
  volume       = {7},
  pages        = {422},
  year         = {2024},
  doi          = {10.1038/s42005-024-01896-z},
}

@book{acrylicstudies,
    author = {Stachiw, J.D.},
    title = {Handbook of Acrylics for Submersibles, Hyperbaric Chambers, and Aquaria},
    publisher = {Best Publishing Company},
    year = {2003}
}

@article{BoniventoArTech,
  author       = {Bonivento, W.M. and Terranova, F.},
  title        = {The science and technology of liquid argon detectors},
  journal      = {Rev. Mod. Phys.},
  volume       = {96},
  pages        = {045001},
  year         = {2024},
  doi          = {10.1103/RevModPhys.96.045001},
}

@article{MonrabalNextWhite,
  author       = {Monrabal, F. and others},
  title        = {The NEXT White (NEW) detector},
  journal      = {Journal of Instrumentation},
  volume       = {13},
  pages        = {P12010},
  year         = {2018},
  doi          = {10.1088/1748-0221/13/12/P12010},
}

@article{CortesiTHGEM,
  author       = {Cortesi, M. and others},
  title        = {Secondary scintillation properties of multi-layer THGEMs operated in low-pressure CF$_4$ and Ar/5\%Xe},
  journal      = {Journal of Instrumentation},
  volume       = {18},
  pages        = {P08005},
  year         = {2023},
  doi          = {10.1088/1748-0221/18/08/P08005},
}

@article{Ban2017,
  author       = {Ban, S. and others},
  title        = {Electroluminescence collection cell as a readout for a high energy resolution Xenon gas TPC},
  journal      = {Nucl. Instr. Meth. A},
  volume       = {875},
  pages        = {185},
  year         = {2017},
  doi          = {10.1016/j.nima.2017.09.015},
}

@article{MavrokoridisGlassGEM,
  author       = {Lowe, A. and Majumdar, K. and Mavrokoridis, K. and Philippou, B. and Roberts, A. and Touramanis, C.},
  title        = {A Novel Manufacturing Process for Glass THGEMs and First Characterisation in an Optical Gaseous Argon TPC},
  journal      = {Appl. Sci.},
  volume       = {11},
  pages        = {9450},
  year         = {2021},
  doi          = {10.3390/app11209450},
}






\end{document}